\documentclass[12pt]{article}
\pdfoutput=1
\usepackage{amsmath}
\usepackage{graphics,graphicx,xcolor}

\def\be{\begin{equation}}
\def\ee{\end{equation}}
\def\arr{\begin{array}{rll}}
\def\ea{\end{array}}
\def\bea{\begin{eqnarray}}
\def\eea{\end{eqnarray}}

\def\N2{$N{=}2$}

\def\>{\rangle}
\def\<{\langle}
\def\+{\dagger}
\def\={\ =\ }

\def\bal{\begin{aligned}}
\def\eal{\end{aligned}}
\begin{document}
\begin{titlepage}
\setcounter{page}{0}
\begin{center}
{\LARGE\bf  Remarks on higher Schwarzians }\\
\vskip 1.5cm
\textrm{\Large Anton Galajinsky \ }
\vskip 0.7cm
{\it
Laboratory of Applied Mathematics and Theoretical Physics, \\
TUSUR, Lenin ave. 40, 634050 Tomsk, Russia} \\

\vskip 0.2cm
{e-mail: a.galajinsky@tusur.ru}
\vskip 0.5cm
\end{center}

\begin{abstract} \noindent
The Schwarzian derivative has recently received renewed attention in connection with the study of the 
Sachdev--Ye--Kitaev model. In mathematics literature, various higher order generalizations of the Schwarzian derivative are known due to Aharonov, Bertilsson, and Schippers.
Physical applications of the higher Schwarzian derivatives have not yet been discussed in any detail. 
In this work, we link Bertilsson's variant to the $\ell$--conformal Galilei group, as well as 
discuss some of its interesting peculiarities. These include a recurrence relation, which allows one to construct the higher Schwarzians iteratively, a composition law, and symmetry transformations.
\end{abstract}

\vspace{0.5cm}

\end{titlepage}
\renewcommand{\thefootnote}{\arabic{footnote}}
\setcounter{footnote}0

\noindent
{\bf 1. Introduction}\\

The Schwarzian derivative (or the Schwarzian, for short) \cite{HS}
\be\label{OS}
S[t'(t);t]=\frac{\frac{d^3 t'}{{d t}^3} }{\frac{d t'}{d t}}-\frac 32 \frac{ {\left(\frac{d^2 t'}{{d t}^2} \right)}^2}{{\left(\frac{d t'}{d t} \right)}^2},
\ee
where $t'(t)$ is a (complex) function of one variable, has gained its rightful place in mathematics by providing a necessary condition for univalence of a function in the unit disc. It had also proved useful for studying geometry of projective lines, 
conformal maps and Teichm\"uller spaces (for a brief review see \cite{OT}).

As far as physical applications are concerned, a remarkable property of the Schwarzian derivative is that it holds invariant under $SL(2,R)$ transformations acting upon the argument
\be\label{SLL} 
t'(t) \quad \to \quad \frac{\alpha t'(t)+\beta}{\gamma t'(t)+\delta},
\ee 
where $\alpha \delta-\beta \gamma=1$. Because $SL(2,R)$ is a finite--dimensional subgroup of the Virasoro group, the Schwarzian derivative arises naturally within the context of string theory and conformal field theory (see, e.g., Chapter 4 in \cite{LT}). More recently, there has been a burst of activity in studying
one--dimensional quantum mechanics that emerges in the low energy limit of a solvable theory displaying maximally chaotic behaviour – the so called Sachdev--Ye--Kitaev model (for a review see \cite{SYK}). A peculiar feature of the system is that its Lagrangian density is proportional to the Schwarzian derivative of a specific function.

Higher order generalizations of (\ref{OS}) are known in mathematics literature \cite{DA,DB,ES}. The first variant was introduced when studying  approximations of an analytic function by M\"obius invariant functions. The so called Aharonov invariants $\phi_n [f(z);z]$, with $n=1,2,\dots$, can be obtained from the recurrence relation \cite{DA}
\be\label{RR}
\frac{d}{dz} \phi_n [f(z);z]=(n+3)  \phi_{n+1} [f(z);z]+\sum_{i=1}^{n-1} \phi_i [f(z);z] \phi_{n-i} [f(z);z],
\ee 
where $\phi_1[f(z);z]=-\frac 16 S[f(z);z]$ and $S[f(z);z]$ is the Schwarzian derivative above. By construction, each $\phi_n [f(z);z]$ holds invariant under $SL(2,R)$ transformations similar to (\ref{SLL}) acting upon the argument $f(z)$.

Another variant of higher Schwarzians was proposed by Bertilsson \cite{DB} who explored estimates for negative powers of the derivative of univalent functions\footnote{Our notation is slightly different from that in \cite{DB}.}
\be\label{BS}
S^{(\frac{n}{2})} [f(z);z]=\frac{2}{n} {\left(\frac{d }{d z} f(z) \right)}^{\frac{n}{2}} \frac{d^{n+1}}{d z^{n+1}}  {\left(\frac{d }{d z} f(z)\right)}^{-\frac{n}{2}},
\ee 
where $n$ is a natural number,
which reproduces $-S[f(z);z]$ at $n=1$. Although $S^{(\frac{n}{2})} [f(z);z]$ is not invariant under a linear fractional change of $f(z)$, it transforms in
a rather peculiar way under such a transformation acting upon the argument of $f(z)$
\be
S^{(\frac{n}{2})} [f(z');z]={\left(\frac{dz'}{dz} \right)}^{n+1} S^{(\frac{n}{2})} [f(z');z'],
\qquad z'=\frac{\alpha z+\beta}{\gamma z+\delta}, 
\ee
with $\alpha \delta-\beta \gamma=1$.
 A similar composition law holds for the original Schwarzian derivative (\ref{OS}) (see Sect. 4).

A third version of higher order Schwarzians $\sigma_n [f(z);z]$, which proved to be invariant under composition on the right with the M\"obius transformation,  was proposed by Schippers in \cite{ES}. Similarly to Aharonov's invariants, these are defined inductively by means of the recurrence relation
\be\label{RRS} 
\sigma_{n+1} [f(z);z]=\frac{d}{dz} \sigma_n [f(z);z]-(n-1) \left(\frac{\frac{d^2 f}{d z^2} }{\frac{d f}{dz} } \right) \sigma_n [f(z);z],
\ee 
where $n \ge 3$ and $\sigma_3 [f(z);z]=S [f(z);z]$.

To the best of our knowledge, physical applications of the higher Schwarzians have not yet been discussed in any detail. The goal of this Letter is to link Bertilsson's variant (\ref{BS}) to the $\ell$--conformal Galilei group \cite{Henkel,NOR}, as well as to discuss some of its interesting peculiarities.

The work is organized as follows. In the next section, we briefly review the $\ell$--conformal Galilei transformations acting in a nonrelativistic spacetime, give generators of infinitesimal transformations, and display the algebra they form. 

In Sect. 3, two physical contexts are discussed, within which the higher order Schwarzians (\ref{BS}) arise naturally. The first example is given by a higher derivative mechanics with the $\ell$--conformal Galilei symmetry \cite{GK}. The second 
instance links to the perfect fluid equations invariant under the $\ell$--conformal Galilei group recently constructed in \cite{AG1,AG2}. 

Peculiar features of the higher Schwarzians (\ref{BS}) are explored in Sect. 4. In particular, a recurrence relation similar to (\ref{RR}) and (\ref{RRS}) is established. A generic composition law is discussed, which links $S^{(\frac{n}{2})} [g(f(z));z]$, where $g(f(z))$ is an arbitrary composite function, to $S^{(\frac{n}{2})} [g(f(z));f(z)]$ and $S^{(\frac{k}{2})} [f(z);z]$ with $k\le n$. Interestingly enough, the composition law for $S^{(\frac{n}{2})} [g(f(z));z]$ with $n>1$ involves the full tower of lower Schwarzians $S^{(\frac{k}{2})} [f(z);z]$, $k\le n$. Furthermore, the law is linear in the Schwarzian derivatives for $n=1,2,3$. A quadratic contribution shows up at $n=4,5,6$. A cubic correction to the composition law arises at $n=7,8,9$. This process appears to continue ad infinitum and the way in which the higher order terms show themselves up resembles a period--doubling bifurcation known from the study of nonlinear phenomena, $n$ being the bifurcation parameter (see e.g \cite{Str}). Symmetries of the higher Schwarzians are discussed as well.

In the concluding Sect. 5, we summarize our results and discuss possible further developments.

In order to illustrate growing complexity in $S^{(\frac n2)} [f(z);z]$ as $n$ increases,
in Appendix we display the derivatives for $n=1,\dots,5$.

Throughout the paper, summation over repeated indices is understood unless otherwise stated.

\vspace{0.5cm}

\noindent
{\bf 2. The $\ell$--conformal Galilei group}\\

Transformations forming the $\ell$--conformal Galilei group include
(temporal) translation, dilatation, and special
conformal transformation, which form $SL(2,R)$ subgroup, as well as spatial rotations, spatial translations, Galilei boosts and constant accelerations \cite{Henkel,NOR}. Given a nonrelativistic spacetime parametrized by a temporal variable $t$ and Cartesian coordinates $x_i$, $i=1,\dots,d$,
a finite form of the transformations reads (no sum over repeated index $n$)
\bea\label{sl2}
&&
t'=\frac{\alpha t+\beta}{\gamma t+\delta}, \qquad
x'_i={\left(\frac{d t'}{d t} \right)}^\ell x_i;
\nonumber\\[2pt]
&&
t'=t, \qquad x'_i=x_i+a^{(n)}_i t^n,
\eea
where $\alpha \delta-\beta \gamma=1$, $\ell$ is an arbitrary (half)integer parameter, and $n=0,\dots, 2\ell$.
Above $n=0$ and $n=1$ correspond to the spatial translation and the Galilei boost, while higher values of $n$ correspond to constant accelerations. The group also contains the conventional $SO(d)$--rotations, $x'_i=x_i+\omega_{ij} x_j$, $\omega_{ij}=-\omega_{ji}$, which in what follows will be disregarded. 

Generators of infinitesimal transformations, which are needed in order to establish structure relations of the corresponding Lie algebra, can be obtained as follows.
Substituting $\alpha=1$, $\delta=1$, $\gamma=0$  into (\ref{sl2}) and regarding $\beta$ as infinitesimal parameter, one obtains the infinitesimal form of the temporal translation
\be
t'=t+\beta, \quad x'_i=x_i.
\ee
Choosing $\alpha=e^{\frac{\lambda}{2}}$, $\delta=e^{-\frac{\lambda}{2}}$, $\beta=0$, $\gamma=0$, setting $\lambda$ to be infinitesimal parameter and Taylor expanding in $\lambda$ up to the first order, one gets the dilatation transformation
\be
t'=t+\lambda t, \qquad x'_i=x_i+\lambda \ell  x_i.
\ee
Infinitesimal form of the special conformal transformation is found by setting $\alpha=1$, $\delta=1$, $\beta=0$, $\gamma=-\sigma$, with infinitesimal $\sigma$, and Taylor expanding in $\sigma$ up to the first order
\be
t'=t+\sigma t^2, \qquad x'_i=x_i+2\sigma \ell t  x_i.
\ee
For infinitesimal parameters $a^{(n)}_i$, with $n=0,\dots, 2\ell$, the acceleration transformations maintain their form (\ref{sl2}).

Introducing the variations $t'=t+\delta t$, $x'_i=x_i+\delta x_i$, and
regarding the operator $\delta$ as the product of an infinitesimal parameter and a generator, one finally obtains the generators
\be
H=\frac{\partial}{\partial t}, \qquad D=t \frac{\partial}{\partial t}+\ell x_i \frac{\partial}{\partial x_i}, \qquad K=t^2 \frac{\partial}{\partial t}+2 \ell t x_i \frac{\partial}{\partial x_i}, \qquad C^{(n)}_i=t^n  \frac{\partial}{\partial x_i},
\ee
which
obey the structure relations of the $\ell$--conformal Galilei algebra \cite{Henkel,NOR}
\begin{align}\label{algebra}
&
[H,D]=H, &&  [H,K]=2 D, && [D,K]=K,
\nonumber\\[2pt]
&
[H,C^{(n)}_i]=n C^{(n-1)}_i, && [D,C^{(n)}_i]=(n-\ell) C^{(n)}_i, && [K,C^{(n)}_i]=(n-2\ell) C^{(n+1)}_i.
\end{align}
Note that, as follows from the last commutator, 
in order to have a finite--dimensional algebra, one has to require $\ell$ to be a half(integer) number.

Recently there has been a burst of activity in studying dynamical realizations of the $\ell$--conformal Galilei group (see e.g. \cite{GK,LSZ,DH2,FIL,DH1,GM,AGKM,AGGM} and references therein). These studies were mainly 
motivated by possible applications within the context of the non–relativistic holography.  

\vspace{0.5cm}

\noindent
{\bf 3. The $\ell$--conformal Galilei group and higher Schwarzian derivatives}\\

We now turn to discuss two physical contexts within which the higher Schwarzian derivatives (\ref{BS}) arise naturally.

As the first example, let us consider an orbit of a nonrelativistic particle moving in $d$ dimensions, which is parametrized by $x_i (t)$, $i=1,\dots,d$. According to our consideration above, 
$SL(2,R)$ subgroup of the $\ell$--conformal Galilei group acts upon the temporal coordinate $t$ and the fields $x_i (t)$ as follows
\be\label{SL2}
t'(t)=\frac{\alpha t+\beta}{\gamma t+\delta}, \qquad
x'_i (t')={\left(\frac{d t'}{d t} \right)}^\ell x_i (t),
\ee
where $\alpha \delta-\beta \gamma=1$, and $\ell$ is a (half)integer parameter. The former relation is now treated as the conformal transformation in one dimension, while the latter condition implies that $x_i (t)$ are primary fields of the conformal dimension $\ell$. 

As far as dynamical realizations of the $\ell$--conformal Galilei group in mechanics are concerned, a free action functional reads \cite{GK}
\be\label{act}
\frac 12 \int dt ~ x_i (t) \frac{d^{n+1} }{d t^{n+1}} x_i (t),
\ee
with $n=2\ell$ and $\ell=\frac 12,1,\frac 32,\dots$. Because for even $n$ (integer $\ell$) the kinetic term turns into a total derivative, only odd values of $n$ (half--integer $\ell$) are usually considered.\footnote{ In two dimensions,  the Levi--Civita symbol $\epsilon_{ij}$ can be used to construct a reasonable kinetic term $\epsilon_{ij} x_i \frac{d^{n+1} }{d t^{n+1}} x_j$ (see e.g. \cite{LSZ}).} A possible interaction potential, which can be added to (\ref{act}), will play no essential role in our subsequent consideration.

When studying specific models with the $\ell$--conformal Galilei symmetry, one usually considers infinitesimal symmetry transformations. However, let us focus on the finite $SL(2,R)$ transformations (\ref{SL2})
and discuss the invariance of (\ref{act}) in more detail. 

At $n=1$ ($\ell=\frac 12$) one finds 
\be 
\frac{d^2}{d t^2} x_i (t)={\left(\frac{d t'}{d t} \right)}^{\frac 32} \frac{d^2}{{d t'}^2} x'_i (t')+\frac 12 {\left(\frac{d t'}{d t} \right)}^{- \frac 12} x'_i (t') ~ S^{(\frac 12)} [t'(t);t],
\ee 
where 
\be\label{ShD}
S^{(\frac 12)} [t'(t);t]=2 {\left(\frac{d t'}{d t} \right)}^{\frac 12} \frac{d^2}{d t^2 } {\left(\frac{d t'}{d t} \right)}^{-\frac 12}=-\left(\frac{\frac{d^3 t'}{{d t}^3} }{\frac{d t'}{d t}}-\frac 32 \frac{ {\left(\frac{d^2 t'}{{d t}^2} \right)}^2}{{\left(\frac{d t'}{d t} \right)}^2}\right).
\ee 
The expression in braces is the Schwarzian derivative (\ref{OS}), which vanishes for  $t'(t)$ in (\ref{SL2}), thus providing the invariance of the action (\ref{act}).

At $n=2$ ($\ell=1$), transformation of the primary field $x_i (t)={\left(\frac{d t'}{d t} \right)}^{-1} x'_i (t')$ results in
\be 
\frac{d^3}{d t^3} x_i (t)={\left(\frac{d t'}{d t} \right)}^{2} \frac{d^3}{{d t'}^3} x'_i (t') +2 \left(  \frac{d}{d t'} x'_i (t') \right) S^{(\frac 12)} [t'(t);t]+{\left(\frac{d t'}{d t} \right)}^{-1} x'_i (t') ~ S^{(1)} [t'(t);t],
\ee 
where we denoted
\be 
S^{(1)} [t'(t);t]={\left(\frac{d t'}{d t} \right)} \frac{d^3}{d t^3 } {\left(\frac{d t'}{d t} \right)}^{-1}=-\left(\frac{\frac{d^4 t'}{{d t}^4} }{\frac{d t'}{d t}}- 6\frac{\left(\frac{d^2 t'}{{d t}^2}\right) \left(\frac{d^3 t'}{{d t}^3}\right)}{{\left(\frac{d t'}{d t}\right)}^2}+ 6\frac{{\left(\frac{d^2 t'}{{d t}^2} \right)}^3}{{\left(\frac{d t'}{d t} \right)}^3} \right).
\ee 
Taking into account the elementary corollaries of (\ref{SL2})
\be\label{suppl}
t=\frac{\beta-\delta t'}{\gamma t'-\alpha}, \qquad  \frac{d t}{d t'}=\frac{1}{{(\gamma t'-\alpha)}^2}, \qquad \frac{d t'}{d t}=\frac{1}{{(\gamma t+\delta)}^2}={(\gamma t'-\alpha)}^2,
\ee
one concludes that  $S^{(1)} [t'(t);t]$ is zero for $t'(t)$ in (\ref{SL2}), which guarantees the symmetry of the action (\ref{act}) at $n=2$.

For higher values of $n=2\ell$, $x_i (t)$ has the conformal dimension $\ell$, 
$x_i (t)={\left(\frac{d t'}{d t} \right)}^{-\ell} x'_i (t')$, and analyzing the way in which
$\frac{d^{n+1} }{d t^{n+1}} x_i (t)$ transforms under  (\ref{SL2}) one 
encounters a higher order generalization of the Schwarzian derivative introduced by Bertilsson in \cite{DB}
\be\label{HS}
S^{(\frac{n}{2})} [t'(t);t]=\frac{2}{n} {\left(\frac{d t'}{d t} \right)}^{\frac{n}{2}} \frac{d^{n+1}}{d t^{n+1}}  {\left(\frac{d t'}{d t} \right)}^{-\frac{n}{2}},
\ee 
where $t'(t)$ is an arbitrary (real or complex) differentiable function of one variable $t$.
The notation, in which both the argument $t'(t)$ and the variable $t$ (with respect to which the Schwarzian derivative is calculated) are explicitly designated,  proves to be particularly helpful for analyzing a composition law (see next section).

As $n$ increases, transformation law of $\frac{d^{n+1} }{d t^{n+1}} x_i (t)$ under (\ref{SL2}) becomes more and more complicated. For $n=1,2,3$ one obtains the relation
\bea\label{lin}
&&
\frac{d^{n+1} }{d t^{n+1}} x_i (t)={\left(\frac{d t'}{d t} \right)}^{\frac{n+2}{2}} \frac{d^{n+1} }{{d t'}^{n+1}} x'_i (t')+\frac{n}{2} {\left(\frac{d t'}{d t} \right)}^{-\frac{n}{2}} x'_i (t') ~ S^{(\frac{n}{2})} [t'(t);t] 
\nonumber\\[2pt]
&&
\qquad \qquad \qquad+\sum_{k=1}^{n-1} {\left(\frac{d t'}{d t} \right)}^{\frac{n}{2}-k} \left(\frac{d^{n-k} }{{d t'}^{n-k}} x'_i (t') \right) \left(\frac{k}{2} C_{n+2}^{k+2} ~ S^{(\frac{k}{2})} [t'(t);t]\right),
\eea
$C_n^k=\frac{n!}{k! (n-k)!}$ being the binomial coefficient, which is linear in the higher Schwarzians. Here and in what follows it is assumed that a sum symbol should be disregarded whenever the upper bound is a negative integer or zero. Also $S^{(\frac{n}{2})} [t'(t);t]$ is set to vanish, if $n$ happens to be a negative integer or zero.

At $n=4$, the expression in braces, which enters the last term in (\ref{lin}), is extended by the quadratic contribution 
\be 
\frac{1}{2^3} \sum_{p=1}^{k-2} p (k-p-1) C_{n+2}^{n-k-1} C_{k+3}^{p+2} ~ S^{(\frac{p}{2})} [t'(t);t] S^{(\frac{k-p-1}{2})} [t'(t);t].
\ee
The resulting formula holds true for $n=5$ and $n=6$ as well, while the cubic correction to it
\bea
&& 
\frac{1}{2^4} \sum_{p,q=1}^{k-4} p q (k-p-q-2) C_{n+2}^{n-k-2} C_{k+3}^{p+q+4} C_{p+q+4}^{3+(p-1)(q-1)(1-\frac 12 (p-2)(q-2))}~ 
\nonumber\\[2pt]
&& 
\qquad \qquad 
\times
S^{(\frac{p}{2})} [t'(t);t] S^{(\frac{q}{2})} [t'(t);t] S^{(\frac{k-p-q-2}{2})} [t'(t);t]
\eea 
shows up at $n=7$. The extended relation continues to be valid for $n=8$ and $n=9$, whereas a quartic contribution of a similar kind arises at $n=10$ and so on and so forth.\footnote{A preliminary consideration suggests that the upper bound of the binomial coefficient $C_{p+q+4}^{3+(p-1)(q-1)(1-\frac 12 (p-2)(q-2))}$ will receive a further correction starting at $n=12$.} Because $S^{(\frac{n}{2})} [\left(\frac{\alpha t+\beta}{\gamma t+\delta} \right);t]=0$ for each value of $n$ (see next section), the invariance of the free action (\ref{act}) under the $\ell$--conformal Galilei transformations (\ref{SL2}) is guaranteed.

The example above demonstrates that the higher Schwarzian derivatives (\ref{HS}) arise naturally within the context of higher derivative mechanics invariant under the  $\ell$--conformal Galilei group. One more example is provided by the perfect fluid equations with the $\ell$--conformal Galilei symmetry \cite{AG1,AG2}
\be\label{fin}
\frac{\partial \rho}{\partial t} + \frac{\partial ( \rho \upsilon_i )}{\partial x_i}=0, \qquad \rho  \mathcal{D}^{2\ell} \upsilon_i=-\frac{\partial p}{\partial x_i},
\ee
with $i=1,\dots,d$. Here $\rho(t,x)$ is the fluid density and $\upsilon_i (t,x)$ is the velocity vector field. $p$ is the pressure obeying the equation of state $p=\nu \rho^{1+\frac{1}{\ell d}}$, in which $\nu$ is a constant. $\mathcal{D}=\frac{\partial}{\partial t} +\upsilon_i  (t,x) \frac{\partial}{\partial x_i}$ is the material derivative. 

Under the action of the $SL(2,R)$ subgroup of the $\ell$--conformal Galilei group the material derivative and the velocity vector field transform as follows (see \cite{AG1} for more details)
\be\label{TRR}
\mathcal{D}=\left(\frac{\partial t'}{\partial t} \right) \mathcal{D}',\qquad
\upsilon_i (t,x)={\left(\frac{\partial t'}{\partial t} \right)}^{1-\ell} \upsilon'_i (t',x')+\frac{\partial}{\partial t} {\left(\frac{\partial t'}{\partial t} \right)}^{-\ell} x'_i,
\ee
which allow one to obtain transformation laws of $\mathcal{D}^n \upsilon_i (t,x)$. For example, at $n=1$ one finds
\be\label{CD}
\mathcal{D} \upsilon_i (t,x)={\left(\frac{\partial t'}{\partial t} \right)}^{2-\ell} \mathcal{D}' \upsilon'_i (t',x')+(1-2\ell) {\left(\frac{\partial t'}{\partial t} \right)}^{-\ell} \left(\frac{\partial^2 t'}{ \partial t^2} \right)  \upsilon'_i (t',x') +\frac{\partial^2}{ \partial t^2} {\left(\frac{\partial t'}{\partial t} \right)}^{-\ell}  x'_i.
\ee
Although for arbitrary value of $\ell$ the field $\mathcal{D} \upsilon_i$ does not transform covariantly, the second term on the right hand side of (\ref{CD}) drops out at $\ell=\frac 12$, while the third term
\be
\frac{\partial^2}{ \partial t^2} {\left(\frac{\partial t'}{\partial t} \right)}^{-\ell}=-\ell  {\left(\frac{\partial t'}{\partial t} \right)}^{-\ell} \left(\frac{\frac{\partial^3 t'}{{\partial t}^3} }{\frac{\partial t'}{\partial t}}-(\ell+1)
\frac{{\left(\frac{\partial^2 t'}{{\partial t}^2} \right)}^2}{{\left(\frac{\partial t'}{\partial t} \right)}^2} \right),
\ee
is proportional to the Schwarzian derivative (\ref{HS}) for $n=1$ ($\ell=\frac 12$). The latter vanishes for the $SL(2,R)$ transformation (\ref{sl2}), thus turning (\ref{CD}) into a covariant expression. Transformation laws of the density and pressure (see \cite{AG1}) then allow one to establish the invariance of the Euler equation $\rho  \mathcal{D} \upsilon_i=-\frac{\partial p}{\partial x_i}$ under the $\ell=\frac 12$ conformal Galilei transformations
(see also the discussion in \cite{JNPP,HZ}).

For $n=2$ one similarly obtains the equality
\bea\label{SCD}
&&
\mathcal{D}^2 \upsilon_i (t,x)={\left(\frac{\partial t'}{\partial t} \right)}^{3-\ell} \mathcal{D}'^2 \upsilon'_i (t',x')
+3(1-\ell) {\left(\frac{\partial t'}{\partial t} \right)}^{1-\ell} \left(\frac{\partial^2 t'}{ \partial t^2} \right) \mathcal{D}' \upsilon'_i (t',x') 
\\[2pt]
&&
\qquad \qquad \quad
+{\left(\frac{\partial t'}{\partial t} \right)}^{1-\ell} \left((1-3\ell)\frac{\frac{\partial^3 t'}{{\partial t}^3} }{\frac{\partial t'}{\partial t}}+3 \ell^2 \frac{{\left(\frac{\partial^2 t'}{{\partial t}^2} \right)}^2}{{\left(\frac{\partial t'}{\partial t} \right)}^2} \right) \upsilon'_i (t',x')
+\frac{\partial^3}{ \partial t^3} {\left(\frac{\partial t'}{\partial t} \right)}^{-\ell}  x'_i,
\nonumber
\eea
which is noncovariant unless $\ell=1$. In the latter case, the second term on the right hand side of (\ref{SCD}) is zero, while the third and forth terms are proportional to $S^{(\frac{1}{2})} [t'(t);t]$ and $S^{(1)} [t'(t);t]$ in (\ref{HS}), respectively, which both vanish for $t'(t)=\frac{\alpha t+\beta}{\gamma t+\delta}$. Transformation properties of the density and pressure (see \cite{AG1}) then guarantee the invariance of the generalized Euler equation $\rho  \mathcal{D}^{2} \upsilon_i=-\frac{\partial p}{\partial x_i}$ under the $\ell=1$ conformal Galilei group. Higher material derivatives of $\upsilon_i (t,x)$ can be treated likewise.

Thus, analyzing the way in which higher material derivatives of the velocity vector field transform under the $SL(2,R)$ subgroup of the $\ell$--conformal Galilei group, one naturally arrives at Bertilsson's variant of the higher Schwarzian derivatives (\ref{HS}). 

\vspace{0.5cm}

\newpage

\noindent
{\bf 4. Properties of the higher Schwarzian derivatives}\\

The higher Schwarzians (\ref{HS}) were introduced in \cite{DB} without discussing their specific properties. In this section, we list some of their interesting peculiarities.

Firstly, as was mentioned in the Introduction, the higher Schwarzians in \cite{DA,ES} are determined by specific recursion relations. For the Schwarzian derivatives in (\ref{HS}) one can similarly obtain the recursive formula
\bea\label{RF}
&& 
S^{(\frac{n+1}{2})} [t'(t);t]=
\frac{d}{d t} S^{(\frac{n}{2})} [t'(t);t]-(n+1)\left(\frac{\frac{d^2 t'}{{d t}^2} }{\frac{d t'}{d t} } \right) S^{(\frac{n}{2})} [t'(t);t] 
\nonumber\\[2pt]
&&
\qquad \qquad \qquad \quad
+\frac 12 \sum_{k=1}^{n-1} C_{n+1}^{k+1} S^{(\frac{k}{2})} [t'(t);t] S^{(\frac{n-k}{2})} [t'(t);t].
\eea 

Secondly, a fundamental property of the original Schwarzian derivative (\ref{OS}) is the composition law
\be\label{CL}
S^{(\frac 12)} [\tau(t'(t));t]=S^{(\frac 12)} [t'(t);t]+{\left(\frac{d t'}{d t} \right)}^2 S^{(\frac 12)} [\tau(t'(t));t'(t)],
\ee
where $\tau(t'(t))$ is an arbitrary composite function. In particular, (\ref{CL}) implies that solving the equation
\be 
S^{(\frac 12)} [\tau(t'(t));t'(t)]=0,
\ee
one can determine a transformation $t'(t) \to \tau(t'(t))$ under which the Schwarzian derivative holds invariant. In this regard, the formula (\ref{HS}) proves to be particularly helpful. The differential equation $\frac{d^{2}}{{d t'}^{2}}  {\left(\frac{d \tau}{d t'} \right)}^{-\frac{1}{2}}=0$ yields
\be 
\tau(t'(t))=\frac{\alpha t'(t)+\beta}{\gamma t'(t)+\delta},
\ee
with $\alpha \delta-\beta \gamma=1$, which is the well known $SL(2,R)$ transformation acting upon the argument of the Schwarzian derivative.

Interestingly enough, a composition law for $S^{(\frac{n}{2})} [\tau(t'(t));t]$ with $n>1$ involves the full tower of lower Schwarzians $S^{(\frac{k}{2})} [t'(t);t]$, $k\le n$. The relation is linear in the Schwarzian derivatives for $n=1,2,3$
\bea\label{LIN} 
&&
S^{(\frac{n}{2})} [\tau(t'(t));t]=S^{(\frac{n}{2})} [t'(t);t]+{\left(\frac{d t'}{d t} \right)}^{n+1} S^{(\frac{n}{2})} [\tau(t'(t));t'(t)]
\\[2pt]
&&
\qquad \qquad \qquad \qquad+\sum_{k=1}^{n-1} {\left(\frac{d t'}{d t} \right)}^{n-k} \left({\left(\frac{d \tau}{d t'} \right)}^{\frac{n}{2}} \frac{d^{n-k}}{ {d t'}^{n-k}}  {\left(\frac{d \tau}{d t'} \right)}^{-\frac{n}{2}}\right) \left( \frac{k}{n} C_{n+2}^{k+2} 
S^{(\frac{k}{2})} [t'(t);t]\right).
\nonumber
\eea
At $n=4$, a quadratic contribution shows up, which extends the last term in braces on the right hand side of (\ref{LIN}) as follows
\be\label{QC}
\frac{k}{n} C_{n+2}^{k+2} 
S^{(\frac{k}{2})} [t'(t);t]+\frac{1}{ 2^2 n} \sum_{p=1}^{k-2} p (k-p-1) C_{n+2}^{n-k-1} C_{k+3}^{p+2} S^{(\frac{p}{2})} [t'(t);t] S^{(\frac{k-p-1}{2})} [t'(t);t].
\ee 
The modified formula holds true for $n=5$ and $n=6$ as well. A cubic correction to (\ref{QC}) appears at $n=7$, which reads
\bea\label{CC}
&&
\frac{1}{2^3 n } \sum_{p,q=1}^{k-4} p q (k-p-q-2) C_{n+2}^{n-k-2} C_{k+3}^{p+q+4} C_{p+q+4}^{3+(p-1)(q-1)(1-\frac 12 (p-2)(q-2))}  
\nonumber\\[2pt]
&&
\qquad 
 \times S^{(\frac{p}{2})} [t'(t);t] S^{(\frac{q}{2})} [t'(t);t] S^{(\frac{k-p-q-2}{2})} [t'(t);t].
\eea 
The resulting expression is valid in the interval $n=1,\dots,9$, whereas at $n=10$ a quartic contribution appears.\footnote{Similarly to our consideration in the preceding section, a preliminary analysis suggests that the upper bound of the binomial coefficient $C_{p+q+4}^{3+(p-1)(q-1)(1-\frac 12 (p-2)(q-2))}$ will receive a further correction starting at $n=12$.} This process appears to continue ad infinitum. Curiously enough, the way in which the higher order terms show themselves up resembles a period--doubling bifurcation known from the study of nonlinear phenomena, $n$ being the bifurcation parameter (see e.g \cite{Str}).

Thirdly, taking into account the definition  (\ref{HS}) and the elementary properties of the linear fractional transformation (\ref{suppl}), one can readily verify that the equality
\be
S^{(\frac{n}{2})} [\left(\frac{\alpha t+\beta}{\gamma t+\delta} \right);t]=0
\ee 
holds for an arbitrary value of $n$.

Finally, because apart from the Schwarzian derivatives themselves the recurrence relation (\ref{RF}) involves the factor $\left(\frac{\frac{d^2 t'}{{d t}^2} }{\frac{d t'}{d t} } \right)$, a symmetry transformation, which is shared by all the higher Schwarzians, appears to be 
\be
t'(t) \quad \to \quad \alpha t'(t)+\beta,
\ee
where $\alpha$ and $\beta$ are arbitrary real parameters.

\vspace{0.5cm}

\noindent
{\bf 5. Conclusion}\\

To summarize, in this work two physical contexts were discussed, within which Bertilsson's variant of the higher Schwarzian derivatives arises naturally. The first example was given by a higher derivative mechanics with the $\ell$--conformal Galilei symmetry \cite{GK}. The second  instance linked to the perfect fluid equations invariant under the $\ell$--conformal Galilei group \cite{AG1,AG2}. Properties of the higher Schwarzians were discussed, which included a recurrence relation, which allowed one to construct the higher Schwarzians iteratively, a composition law, and symmetry transformations.

Turning to possible further developments, it would be interesting to study whether the higher order generalizations of the Schwarzian derivative in \cite{DA,DB,ES} might prove useful for holographic applications. 

As was mentioned above, the way in which the higher order terms enter the composition law resembles a period--doubling bifurcation, $n$ being the bifurcation parameter.
A possible connection of the higher Schwarzians to nonlinear phenomena deserves a separate study.

In a recent work \cite{AG3}, a group--theoretic approach to the construction of the  Schwarzian derivative (\ref{OS}) was proposed (for further developments see \cite{KK} and references therein). It would be interesting to understand whether 
the higher order variants in \cite{DA,DB,ES} can be constructed in a similar way. 

Possible supersymmetric generalizations of the higher Schwarzians in \cite{DA,DB,ES} are worth exploring as well.

\vspace{0.5cm}

\noindent{\bf Acknowledgements}\\

This work was supported by the Russian Science Foundation, grant No 23-11-00002.

\vspace{0.5cm}

\noindent{\bf Appendix}\\

In order to illustrate growing complexity in $S^{(\frac n2)} [t'(t);t]$ as $n$ increases,
in this Appendix we display the derivatives for $n=1,\dots,5$ 
\bea
&&
S^{(\frac 12)} [t'(t);t]=-\left(\frac{\frac{d^3 t'}{{d t}^3} }{\frac{d t'}{d t}}-\frac 32 \frac{ {\left(\frac{d^2 t'}{{d t}^2} \right)}^2}{{\left(\frac{d t'}{d t} \right)}^2}\right),
\nonumber\\[2pt]
&&
S^{(1)} [t'(t);t]=-\left(\frac{\frac{d^4 t'}{{d t}^4} }{\frac{d t'}{d t}}- 6\frac{\left(\frac{d^2 t'}{{d t}^2}\right) \left(\frac{d^3 t'}{{d t}^3}\right)}{{\left(\frac{d t'}{d t}\right)}^2}+ 6\frac{{\left(\frac{d^2 t'}{{d t}^2} \right)}^3}{{\left(\frac{d t'}{d t} \right)}^3} \right),
\nonumber\\[2pt]
&&
S^{(\frac 32)} [t'(t);t]=-\left(\frac{\frac{d^5 t'}{{d t}^5} }{\frac{d t'}{d t}} -10 \frac{ \left(\frac{d^2 t'}{{d t}^2} \right) \left(\frac{d^4 t'}{{d t}^4} \right)} {{\left(\frac{d t'}{d t} \right)}^2}-\frac{15}{2} \frac{{\left(\frac{d^3 t'}{{d t}^3} \right)}^2}{{\left(\frac{d t'}{d t} \right)}^2} +\frac{105}{2} \frac{{\left(\frac{d^2 t'}{{d t}^2} \right)}^2 \left(\frac{d^3 t'}{{d t}^3} \right)}{{\left(\frac{d t'}{d t} \right)}^3}-\frac{315}{8} \frac{{\left(\frac{d^2 t'}{{d t}^2} \right)}^4 }{{\left(\frac{d t'}{d t} \right)}^4}\right),
\nonumber\\[2pt]
&&
S^{(2)} [t'(t);t]=-\left(\frac{\frac{d^6 t'}{{d t}^6} }{\frac{d t'}{d t}}-15 \frac{\left(\frac{d^2 t'}{{d t}^2} \right) \left(\frac{d^5 t'}{{d t}^5} \right) }{{\left(\frac{d t'}{d t} \right)}^2} -30 \frac{\left(\frac{d^3 t'}{{d t}^3} \right) \left(\frac{d^4 t'}{{d t}^4} \right) }{{\left(\frac{d t'}{d t} \right)}^2}+120 \frac{{\left(\frac{d^2 t'}{{d t}^2} \right)}^2 \left(\frac{d^4 t'}{{d t}^4} \right) }{{\left(\frac{d t'}{d t} \right)}^3}
\right.
\nonumber\\[2pt]
&&
\left.
\qquad \qquad \qquad 
+180 \frac{\left(\frac{d^2 t'}{{d t}^2} \right) {\left(\frac{d^3 t'}{{d t}^3} \right)}^2 }{{\left(\frac{d t'}{d t} \right)}^3}
-600 \frac{{\left(\frac{d^2 t'}{{d t}^2} \right)}^3 \left(\frac{d^3 t'}{{d t}^3} \right)}{{\left(\frac{d t'}{d t} \right)}^4}+360  \frac{{\left(\frac{d^2 t'}{{d t}^2} \right)}^5  }{{\left(\frac{d t'}{d t} \right)}^5} \right),
\nonumber\\[2pt]
&&
S^{(\frac 52)} [t'(t);t]=-\left( 
\frac{\frac{d^7 t'}{{d t}^7} }{\frac{d t'}{d t}}-21 \frac{\left(\frac{d^2 t'}{{d t}^2} \right) \left(\frac{d^6 t'}{{d t}^6} \right) }{{\left(\frac{d t'}{d t} \right)}^2} -\frac{105}{2} \frac{\left(\frac{d^3 t'}{{d t}^3} \right) \left(\frac{d^5 t'}{{d t}^5} \right) }{{\left(\frac{d t'}{d t} \right)}^2}
+\frac{945}{4} \frac{{\left(\frac{d^2 t'}{{d t}^2} \right)}^2 \left(\frac{d^5 t'}{{d t}^5} \right) }{{\left(\frac{d t'}{d t} \right)}^3}
\right.
\nonumber
\eea 
\bea
&&
\left.
\qquad \qquad \qquad 
-35 \frac{{\left(\frac{d^4 t'}{{d t}^4} \right)}^2 }{{\left(\frac{d t'}{d t} \right)}^2}
+945 \frac{ \left(\frac{d^2 t'}{{d t}^2} \right) \left(\frac{d^3 t'}{{d t}^3} \right)\left(\frac{d^4 t'}{{d t}^4} \right)}{{\left(\frac{d t'}{d t} \right)}^3}-\frac{3465}{2} \frac{{\left(\frac{d^2 t'}{{d t}^2} \right)}^3 \left(\frac{d^4 t'}{{d t}^4} \right) }{{\left(\frac{d t'}{d t} \right)}^4} +\frac{945}{4} \frac{{\left(\frac{d^3 t'}{{d t}^3} \right)}^3 }{{\left(\frac{d t'}{d t} \right)}^3}
\right.
\nonumber\\[2pt]
&&
\left.
\qquad \qquad \qquad 
-\frac{31185}{8} \frac{ {\left(\frac{d^2 t'}{{d t}^2} \right)}^2 {\left(\frac{d^3 t'}{{d t}^3} \right)}^2 }{{\left(\frac{d t'}{d t} \right)}^4}+\frac{135135}{16} \frac{{\left(\frac{d^2 t'}{{d t}^2} \right)}^4 \left(\frac{d^3 t'}{{d t}^3} \right) }{{\left(\frac{d t'}{d t} \right)}^5}-\frac{135135}{32} \frac{{\left(\frac{d^2 t'}{{d t}^2} \right)}^6  }{{\left(\frac{d t'}{d t} \right)}^6} \right).
\nonumber
\eea

\end{document}